\documentclass[manuscript,review]{acmart}

\newcommand{\workshopname}{GenAICHI: CHI 2025 Workshop on Generative AI and HCI}
\newcommand{\licensedetails}{Licensed under a Creative Commons Attribution 4.0 International License (CC BY 4.0). Copyright remains with the author(s).}
\newcommand\extrafootertext[1]{
    \bgroup
    \renewcommand\thefootnote{\fnsymbol{footnote}}%
    \renewcommand\thempfootnote{\fnsymbol{mpfootnote}}%
    \footnotetext[0]{#1}%
    \egroup
}

\usepackage[autostyle, english = american]{csquotes}
\MakeOuterQuote{"}
\usepackage{hyperref}
\usepackage{cleveref}
\usepackage{caption}

\AtBeginDocument{ 
    \fancypagestyle{firstpagestyle}{
        \fancyhf{}
        \fancyfoot[L]{\sffamily\footnotesize \workshopname}%
        \fancyfoot[C]{\sffamily\footnotesize \thepage}
    }
    \fancyhf{}
    \fancyhead[L]{\sffamily\footnotesize\shorttitle}
    \fancyhead[R]{\sffamily\footnotesize\shortauthors}
    \fancyfoot[L]{\sffamily\footnotesize\workshopname}%
    \fancyfoot[C]{\sffamily\footnotesize\thepage}
    \extrafootertext{\licensedetails}
}
\usepackage{lipsum} 

\usepackage{xr}

\begin{document}

\title{From Teacher to Colleague: How Coding Experience Shapes Developer Perceptions of AI Tools}

\author{Ilya Zakharov}
\email{ilia.zaharov@jetbrains.com}
\affiliation{%
  \institution{JetBrains Research}
  \city{Belgrade}
  \country{Serbia}
}

\author{Ekaterina Koshchenko}
\email{ekaterina.koshchenko@jetbrains.com}
\affiliation{%
  \institution{JetBrains Research}
  \city{Amsterdam}
  \country{Netherlands}
}

\author{Agnia Sergeyuk}
\email{agnia.sergeyuk@jetbrains.com}
\affiliation{%
    \institution{JetBrains Research}
    \city{Belgrade}
    \country{Serbia}
}

\begin{abstract}

AI-assisted development tools promise productivity gains and improved code quality, yet their adoption among developers remains inconsistent. Prior research suggests that professional expertise influences technology adoption, but its role in shaping developers’ perceptions of AI tools is unclear. We analyze survey data from 3380 developers to examine how coding experience relates to AI awareness, adoption, and the roles developers assign to AI in their workflow. Our findings reveal that coding experience does not predict AI adoption but significantly influences mental models of AI’s role. Experienced developers are more likely to perceive AI as a junior colleague, a content generator, or assign it no role, whereas less experienced developers primarily view AI as a teacher. These insights suggest that AI tools must align with developers’ expertise levels to drive meaningful adoption.

\end{abstract}


\maketitle

\section{Introduction}

AI-assisted development tools are increasingly being integrated into professional coding environments, promising productivity gains, code quality improvements, and automation of routine tasks \cite{vasiliniuc2023case,imai2022github,peng2023impact}. Despite these advancements, AI adoption varies widely among developers, with some readily integrating AI-driven tools into their workflows while others remain skeptical or resistant. Prior research suggests that personal experience and technical knowledge shape users' mental models, which in turn influence their perception of a tool's usability and affect adoption \cite{chen2018personal}, suggesting that the more experienced professionals are, the more they are prone to the adoption of the new tools in the field. However, it is unclear whether this trend holds in software engineering and what ``roles''  --- specific functions AI can perform within a developer's workflow, representing mental models of a tool --- are associated with its adoption.

Our research seeks to understand the connection between professional expertise, in our case in coding, and individuals' perspectives on AI. Overall, our research questions may be formulated as follows:
\begin{itemize}
    \item \textbf{RQ1}: \textit{How is coding experience related to the adoption of AI tools?}
    \item \textbf{RQ2}: \textit{How do coding experience and exposure to AI tools relate to the ``roles'' that developers can assign for AI in their workflow?}
\end{itemize}

In analyzing survey data from 3380 developers conducted in Q2 2024~\cite{JetBrains2024}, we found that developers’ experience does not predict whether they adopt or are even aware of AI tools. However, experience does shape the roles they assign to these tools. Specifically, more experienced developers are more likely to perceive AI as either a junior colleague, a content generator, or assign it no role at all. In contrast, less experienced developers tend to describe AI in a teacher role. These findings suggest that senior developers may not perceive enough value in AI tooling to integrate it into their workflow, whereas novice programmers may rely on AI tools for guidance and learning. This highlights the importance of aligning AI tools with users’ mental models --- especially for developers with varying levels of expertise --- by designing features that address their distinct expectations and needs.

\section{Related work}

The relationship between professional expertise and the adoption of AI tools in software development is complex. Drawing on established models of technology acceptance, prior research indicates that external variables --- such as organizational context and prior experience --- play a pivotal role in shaping attitudes toward new technologies~\cite{venkatesh2003user,davis1989perceived}. 

Prior research suggests that personal experience and technical knowledge shape users’ mental models, which in turn influence their perception of a tool’s usability and affect adoption \cite{chen2018personal}. A recent study on how developers perceive and interact with AI-driven code completion suggests that discrepancies between a developer's mental model and the tool's operations can lead to resistance to its adoption~\cite{desolda2025understanding}. Mental model mismatches have been shown to result in frustration, reduced trust, and eventual abandonment of AI tools~\cite{dhuliawala2023diachronic}.

Based on \citet{ericsson2002prospects} general theory of expertise, it might be suggested that experts tend to rely on years of deliberate practice and accumulated domain knowledge, thereby requiring robust evidence of possible benefits before adopting AI systems. In contrast, novices are often more open to experimentation and may rapidly complete tasks with AI tools. However, they tend to over-rely on it, potentially veering away from correct solutions, lacking the expertise to catch subtle errors \cite{prather2023s}, sometimes relying on surface-level cues of the tool functionality~\cite{yarlas2000problem}.

Prior research offers mixed findings on how expertise influences AI adoption, making its impact on software developers unclear. Moreover, while mental models shape adoption, little is known about how experience levels affect the "roles" developers assign to AI tools. We think that understanding this interplay is key to designing AI tools that align with developers' needs and encourage broader adoption.

\section{Method}

To study the relationship between coding experience and AI tool adoption, as well as how developers' mental models shape the "roles" they assign to AI within their workflow, we analyzed the subset of data from JetBrains' State of Developer Ecosystem Report 2024~\cite{JetBrains2024}. This report is designed to track developers' habits and attitudes towards various programming languages, applications, tools, frameworks, and even developers' lifestyles and habits. The main survey was conducted in May–June 2024, and the main report is based on the input of 23,262 developers from 171 countries and regions. We used a subset of the whole survey that relates to the study's research questions. The overall methodology for the survey can be found through the link~\footnote{JetBrains' State of Developer Ecosystem Report Methodology --- \url{https://www.jetbrains.com/lp/devecosystem-2024/methodology/}}. 

For the purposes of the present research, we derived and analyzed answers to the following questions regarding:
\begin{itemize}
    \item Participants' coding experience in years;
    \item The total number of AI tools for development people tried ("AI experience") and heard about ("AI awareness");
    \item Roles of AI tools for developers.
\end{itemize}
The roles suggested in the survey were based on prior studies in the field~\cite{ross2023programmer} and previous surveys conducted by JetBrains~\footnote{JetBrains' State of Developer Ecosystem Reports --- \url{https://blog.jetbrains.com/category/deveco/}}. The available options included: assistant, tool, reference guide, content generator, problem solver, collaborator, senior colleague, junior colleague, teacher, advisor, reviewer, copilot, and companion. Additionally, there was an option to specify your own role or select "None of the above".

The acquired subset was cleaned to focus on developers (people who indicated "Coding/Programming" as a part of their work activities), reducing the initial sample to 19820 participants. From this sample,  8073 participants answered the questions related to their experience with AI, and 3380 participants answered chose at least one option for the question "In which roles do you see AI tools for developers".

To understand the relationship between coding experience and adoption of AI tools, we calculated Spearman's correlations between coding experience, "AI experience," and "AI awareness" metrics. To examine the association between coding experience and the likelihood of a respondent selecting a given role, we fitted a series of logistic regression models, estimated using the maximum likelihood method. The estimated coefficient for coding experience was exponentiated to yield an odds ratio, which quantifies the change in odds of selecting a given role for each additional unit of coding experience. P-values were adjusted for the multiple comparisons using Benjamini \& Hochberg false discovery rate procedure \cite{benjamini_controlling_1995}.

\section{Results}

\begin{table}[hb]
\centering
\begin{tabular}{lccc}
\toprule
\textbf{Variables} & \textbf{Spearman's $\rho$} & \textbf{p-value} & \textbf{N} \\
\midrule
\textbf{AI Experience vs. AI Awareness}    & 0.546 & $<0.001$ & 8073 \\
\textbf{AI Awareness vs. Coding Experience} & 0.031 & 0.005 & 8073 \\
AI Experience vs. Coding Experience   & -0.005 & 0.652 & 8073 \\
\bottomrule
\end{tabular}
\captionsetup{justification=centering}
\caption{Spearman Correlation Results Among AI Awareness, AI Experience, and Coding Experience \newline
Significant correlations are shown in bold.}
\label{tab:corrs}
\end{table}

\begin{table}[ht]
\centering
\begin{tabular}{r r r r r r r r r r r}
\toprule
Role & {Coefficient} & {Std. Err.} & {z} & {P$>$|z|} & {Odds Ratio} & {CI Lower} & {CI Upper} & {p-adj} \\
\midrule
\textbf{None of the above }               & 0.322 & 0.080 & 4.036 & 0.000 & 1.380 & 1.180 & 1.614 & 0.001 \\
\textbf{As a teacher}                     & -0.124 & 0.042 & -2.986 & 0.003 & 0.883 & 0.814 & 0.958 & 0.017 \\
\textbf{As a junior colleague}            & 0.157 & 0.053 & 2.935 & 0.003 & 1.170 & 1.053 & 1.299 & 0.017 \\
\textbf{As a content generator}           & 0.083 & 0.030 & 2.780 & 0.005 & 1.087 & 1.025 & 1.152 & 0.020 \\
As a senior colleague            & -0.134 & 0.074 & -1.803 & 0.071 & 0.875 & 0.757 & 1.012 & 0.214 \\
As a problem solver              & -0.034 & 0.029 & -1.163 & 0.245 & 0.967 & 0.913 & 1.023 & 0.513 \\
As a copilot                     & 0.027 & 0.024 & 1.095 & 0.274 & 1.027 & 0.979 & 1.078 & 0.513 \\
As an advisor                   & 0.030 & 0.029 & 1.017 & 0.309 & 1.030 & 0.973 & 1.092 & 0.515 \\
As a reference guide             & 0.014 & 0.023 & 0.610 & 0.542 & 1.014 & 0.970 & 1.060 & 0.812 \\
As a tool                        & -0.009 & 0.019 & -0.459 & 0.646 & 0.991 & 0.956 & 1.028 & 0.875 \\
As a collaborator                & 0.016 & 0.041 & 0.386 & 0.700 & 1.016 & 0.937 & 1.101 & 0.875 \\
As an assistant                  & -0.004 & 0.019 & -0.207 & 0.836 & 0.996 & 0.960 & 1.034 & 0.919 \\
As a reviewer                   & 0.005 & 0.030 & 0.180 & 0.857 & 1.005 & 0.948 & 1.066 & 0.919 \\
As a companion                  & -0.003 & 0.037 & -0.070 & 0.944 & 0.997 & 0.928 & 1.072 & 0.944 \\
Other, please specify:           & 0.176 & 0.111 & 1.578 & 0.114 & 1.192 & 0.958 & 1.483 & 0.286 \\

\bottomrule
\end{tabular}
\captionsetup{justification=centering}

\caption{Logistic regression results for AI roles. The ``p-adj'' represents the FDR-adjusted p-value.\newline
Significant associations are shown in bold.}
\label{tab:ai_roles_results}
\end{table}

We first examined the relationship between Coding Experience and both Awareness of and Experience with AI tools for coding. As shown in \Cref{tab:corrs}, AI Awareness and AI Experience are significantly correlated, as expected. However, there is no correlation between Coding Experience and AI Experience, and the correlation between Coding Experience and AI Awareness, while statistically significant, is negligible (r=0.031).

Next, we examined whether Coding Experience is associated with the specific "roles" users assign to AI tools. Logistic regression results (\Cref{tab:ai_roles_results}) show that four response types are significantly linked to Coding Experience: three specific roles --- "teacher", "junior colleague", and "content generator" --- as well as the "None of the above" option. The odds ratios indicate that developers with more coding experience are more likely to choose "junior colleague", "content generator", or "None of the above", while those with less experience more often select "teacher".

\section{Discussion and Conclusion}

The goal of our study was to investigate the relationship between Coding Experience, Awareness of and Experience with AI tools for development, and perceptions of them. 

Although the correlation analysis indicated a statistically significant relationship between coding experience and AI Awareness, the extremely low correlation coefficient (r = 0.031) suggests that coding experience alone accounts for only a negligible portion of the variance in AI Awareness. This finding implies that other factors ---such as exposure to technology, organizational culture, or individual curiosity --- might be more influential in shaping awareness of AI tools.

The logistic regression analysis demonstrates that coding experience is associated with distinct "roles" that users assign to AI tools. Participants with higher levels of coding experience tend to perceive the AI tool as a junior colleague or content generator. Conversely, less experienced coders tended to view the AI tool as a teacher. This divergence may reflect underlying differences in expectations: novice programmers might seek instructional support and guidance from AI systems, whereas more seasoned developers may be looking for a collaborative but not necessarily reliable partner that can help with content generation and problem solving.

Notably, developers with more coding experience also chose the "None of the above" option significantly more often. This may suggest that experienced professionals might be relatively rigid or skeptical toward new emerging technologies and that a considerable part of that community currently tends not to see any useful "role" for AI tools for coding. This observation is particularly interesting given the negligible association between coding expertise and AI Awareness and the association between expertise and AI Experience. In other words, individuals with greater coding expertise do not seem to demonstrate a higher interest in experimenting with AI tools for coding.

To sum up, our findings indicate that while coding experience minimally impacts the adoption of AI tools, it significantly shapes developers’ perceptions of the "role" of AI tooling in their workflow and their willingness to engage with such tools.

\begin{acks}
We thank the JetBrains Market Research \& Analytics Team, especially the Surveys Team, for designing and conducting the State of Developer Ecosystem Survey and for their support throughout our study.
\end{acks}

\section*{Data and Code Availability Statement}

The JetBrains' State of Developer Ecosystem Report 2024 is available at ~\cite{JetBrains2024}. The code for data cleaning and analysis is available upon request.

\bibliographystyle{ACM-Reference-Format}
\bibliography{references}

\end{document}